\documentclass{PoS}
\pdfoutput=1
\usepackage{url}
\usepackage{siunitx}

\title{Following up Transient Sources at Very High Energies with MAGIC}

\ShortTitle{Following up Transient Sources at Very High Energies with MAGIC}

\author{\speaker{Alessio Berti}$^{1}$, Elisa Bernardini$^{2}$, Wrijupan Bhattacharyya$^{2}$, Juan Cortina$^{3}$, Stefano Covino$^{4}$, Daniela Dorner$^{5}$, Elia do Souto Espi\~neira$^{6}$, Alicia Fattorini$^{7}$, Luca Foffano$^{8}$, Satoshi Fukami$^{9}$, Markus Garczarczyk$^{2}$, John Hoang$^{10}$, Susumu Inoue$^{11}$, Francesco Longo$^{12}$, Marina Manganaro$^{13}$, Davide Miceli$^{14}$, Elena Moretti$^{6}$, Koji Noda$^{9}$, Michele Peresano$^{14}$, Marc Rib\'o$^{15}$, Konstancja Satalecka$^{2}$, Antonio Stamerra$^{16}$, Yusuke Suda$^{17}$, Martin Will$^{17}$ for the MAGIC collaboration\footnote{\texttt{https://magic.mpp.mpg.de/}. For collaboration list see PoS(ICRC2019)1177.}\\
$^{1}$University of Torino and INFN Torino, Torino, Italy \\
$^{2}$Deutsches Elektronen-Synchrotron (DESY), Zeuthen, Germany \\
$^{3}$CIEMAT, Madrid, Spain $^{4}$INAF, Osservatorio Astronomico di Brera, Merate, Italy \\
$^{5}$Universit\"at W\"urzburg, W\"urzburg, Germany $^{6}$IFAE-BIST, Bellaterra (Barcelona), Spain \\
$^{7}$Technische Universit\"at Dortmund, Dortmund, Germany $^{8}$University of Padova and INFN Padova, Padova, Italy $^{9}$ICRR, University of Tokyo, Kashiwa, Chiba, Japan \\
$^{10}$Instituto de Part\'iculas y Cosmolog\'ia (IPARCOS), Universidad Complutense, Madrid, Spain \\
$^{11}$RIKEN, Wako, Saitama, Japan $^{12}$University of Trieste and INFN Trieste, Trieste, Italy \\
$^{13}$University of Rijeka, Rijeka, Croatia $^{14}$University of Udine and INFN Trieste, Udine, Italy \\
$^{15}$Universitat de Barcelona, ICCUB, IEEC-UB, Barcelona, Spain $^{16}$INAF, National Institute for Astrophysics, Rome, Italy $^{17}$Max-Planck-Institut f\"ur Physik, Munich, Germany \\
E-mail: \email{Alessio.Berti@to.infn.it}}

\abstract{Several classes of sources are known to emit different messengers. Among them, transient sources are a special case, due to their serendipitous occurrence, time variability and duration on different timescales. They are associated with explosive and catastrophic events where very compact objects like neutron stars and black holes are involved. The difficulty of observing such elusive and possibly short-lasting events requires a fast reaction and a well-organized alert network between different experiments. In order to characterize them in the best possible way, instruments with a wide field of view should serve as external triggers for facilities with small sky coverage. MAGIC, as a Cherenkov telescope, belongs to the latter category. The search for transients by MAGIC is possible thanks to an automatic alert system listening to the alerts sent by the Gamma-ray Coordinate Network (GCN). In this contribution we describe the MAGIC alert system, which was designed mainly for the follow-up of Gamma-Ray Bursts in its initial conception. The alert system was recently updated in a multi-messenger context, receiving alerts also from neutrino and GW observatories. Finally we will present the MAGIC program for transient sources and how it was adapted in the current multi-wavelength and multi-messenger panorama.}

\FullConference{36th International Cosmic Ray Conference -ICRC2019-\\
		July 24th - August 1st, 2019\\
		Madison, WI, U.S.A.}

\begin{document}

\section{Introduction}
In the era of multi-messenger astrophysics, the observation of transient sources is becoming more and more crucial. These events, considering their connection with extreme objects and environments, emit several distinct messengers. The coordination of different experiments is then crucial to understand their nature. In the recent years, with the full operation of IceCube and the improved sensitivity of gravitational-wave (GW) ground-based interferometers (LIGO and Virgo), the search of a (temporally and/or spatially) correlated $\gamma$-ray flux from transient events became fundamental: it could help to find the answer to some long-standing questions like the origin of ultra high energy cosmic rays and of high energy astrophysical neutrinos. Indeed, some of the candidate sources which are thought to produce high energy neutrinos and/or gravitational waves are known to be $\gamma$-ray emitters or are supposed to be one. Furthermore, some of these classes of sources are transients (e.g.\ Gamma-Ray Bursts) or may display a transient behavior (e.g.\ flares from Active Galactic Nuclei). Therefore, it is of utterly importance to observe transient events in a broad range of the electromagnetic spectrum, for example in the very high energy band (VHE, $\gamma$-rays with $E\gtrsim\SI{100}{\GeV}$). In this energy range, up to few tens of \si{\TeV}, transient searches are performed with ground-based experiments like Imaging Atmospheric Cherenkov Telescopes (IACTs). Given the intrinsic serendipitous nature of transient events, performing their follow-up constitutes a hard challenge for IACTs, since they need to be triggered and react as fast as possible. \\
In this contribution we will focus on the rich transient program of MAGIC, one of the currently operating Cherenkov telescope systems, and describe the solutions adopted to perform the follow-up of this elusive class of astrophysical sources.

\section{The MAGIC Telescopes}
MAGIC\footnote{\url{https://magic.mpp.mpg.de/}} (\textit{Major Atmospheric Gamma Imaging Cherenkov}) is a system of two Cherenkov telescopes located at \SI{2200}{\meter} a.s.l. within the Observatorio Roque de Los Muchachos (ORM) in the Canary island of La Palma, Spain (\ang{28.8} N, \ang{17.9} W) \cite{Aleksicetal2016a}. Each telescope has a light-weight fiber carbon structure which supports the reflector and the camera. The former has a \SI{17}{\meter} diameter surface, while the latter hosts 1039 photomultipliers of 1 inch each. The large reflective surface gives MAGIC the possibility to detect gamma-rays from about $\SI{50}{\GeV}$ (at zenith, trigger level), allowing for a good energy overlap with space-based gamma instruments like \textit{Fermi}-LAT and AGILE. MAGIC started its operations in 2004 as a single telescope and from 2009 as a stereo system with the addition of the second telescope. Moreover, the telescopes underwent a major upgrade in 2011-2012: in particular the camera of MAGIC-I was replaced by a clone of the one of MAGIC-II and the digitization electronics was upgraded. These technical upgrades, accompanied also by a boost in the analysis methods, led to the current performance of the MAGIC system, summarized in the following \cite{Aleksicetal2016b}: an integral sensitivity as low as 0.7\% of the Crab Nebula flux above \SI{220}{\GeV} in \SI{50}{\hour}; an angular resolution of $\sim\ang{0.06}$ (\ang{0.1}) at \SI{1}{\TeV} (\SI{100}{\GeV}) and an energy resolution of 15\% (24\%) at \SI{1}{\TeV} (\SI{100}{\GeV}). \\
Thanks to these figures of merit and the fast slewing capabilities ($\ang{7}$/s in \textit{fast mode}), MAGIC is a very suitable to perform the follow-up of transient source in the VHE range.

\section{The MAGIC Transients Program}

The design of the MAGIC telescopes was from the beginning focused on fast repositioning and low energy threshold in order to catch the possible VHE emission from Gamma-Ray Bursts (GRBs). In the last years, the MAGIC Transient scientific program has been extended to include the follow-up of alerts coming from facilities detecting neutrinos (e.g.\ IceCube) and gravitational waves (e.g.\ LIGO and Virgo). Recently, a new class of transient sources, Fast Radio Bursts (FRBs), has been discovered and included in the MAGIC Transient program. Clearly, the fraction of observational time dedicated to scientific projects related to transient searches is noticeable, considering also the world-wide effort towards multi-wavelength and multi-messenger astrophysics.

\subsection{The MAGIC Automatic Alert System}
MAGIC adopted several solutions in order to perform the follow-up of transient events. The major limitations of IACTs are the low duty cycle and the small field of view (\ang{3.5} for MAGIC). In the case of MAGIC the maximum duty cycle can be increased up to $\sim$40\% by performing observation under low and moderate moonlight thanks to specific hardware configurations \cite{Ahnenetal2017}. The small field of view prevents MAGIC to observe transient sources without the help of external triggers: only instruments with a large field of view can provide the localization of transient events to IACTs. For this reason, MAGIC developed an Automatic Alert System with the following tasks: (a) receive alerts from the \textit{Gamma-ray Coordinate Network} (GCN), which disseminates information about transient events from space-born and ground-based facilities to interested partners; (b) process the alerts according to predefined criteria and check if the target is observable from the MAGIC site; (c) in the cases it is foreseen, communicate with the Central Control of the MAGIC telescope to start the automatic transient alert follow-up procedure. \\
The aim of such procedure is to activate the fast slewing of the telescopes to the target position and to prepare them for data taking (DAQ initialization, adjustment and focusing of mirrors, trigger and thresholds setting). In its initial concept, the alert system was designed to respond only to GRB alerts generated by satellites like \textit{Fermi}, \textit{Swift} and \textit{INTEGRAL}. As GCN started to distribute alerts coming from neutrino and GW detectors, the MAGIC AAS was adapted in order to receive and process these new alert streams. In this way the system gives the possibility of contributing to multi-messenger observations.

\subsection{Gamma-Ray Bursts}
GRBs are most probably the archetype of transient sources. They appear as serendipitous flashes of gamma rays in random positions in the sky. Their energy release up to $\sim\num{e52}-\SI{e53}{erg}$ (if isotropic), happening in a short time, makes them the most energetic sources of electromagnetic radiation. Most of the energy is released in the so called \textit{prompt} phase of the GRB, which is followed by a fainter emission called \textit{afterglow}, which can be observed over a wide energy range along the electromagnetic spectrum, from radio to high energy gamma rays. \\
Several theoretical models predict that GRBs should be capable of producing VHE gamma rays detectable by current IACTs. The challenges of GRB follow-up by IACT are mostly related to their sudden appearance and fast fading. In the case of MAGIC, this issue is tackled thanks to the aforementioned AAS and the fast slewing. Another issue is the typical far distance of GRBs which increases the probability of gamma-ray absorption with the Extragalactic Background Light (EBL). The effect of the EBL increases with redshift and with the energy of gamma-rays: the low threshold of MAGIC allows to observe distant sources in an energy range which is less (or not) affected by EBL absorption. \\
The detection of VHE emission from GRBs has always been one of the primary goals of the MAGIC experiment. Up to January 2019, in more than fifteen years of operation, MAGIC performed the follow-up of more than 100 GRBs in favorable conditions (good weather and without technical issues). Owing to this unceasing search, on 14th January 2019 MAGIC detected for the first time VHE gamma rays above \SI{300}{\GeV} from GRB\,190114C \cite{Mirzoyanetal2019a,Mirzoyanetal2019b}, a long GRB detected by \textit{Fermi}-GBM and \textit{Swift}-BAT. Even if the observation was performed with moderate moonlight conditions and at rather high zenith angle, the significance of the signal is high, more than 20 standard deviations in the first 20 minutes of data. Such detection will have strong implications on GRB physics, in particular on the understanding of the underlying physical process generating the VHE emission, constraining or possibly ruling out theoretical models. Further implications outside GRB physics are expected regarding EBL studies (the GRB is at redshift $z=0.4245$) and possibly physics beyond the standard model (e.g.\ Lorentz Invariance Violation). \\
Another event of interest in the MAGIC GRB sample is the short and close ($z=0.16$) GRB\,160821B, recently associated to a kilonova \cite{Lambetal2019,Trojaetal2019}. In the MAGIC data the GRB shows a hint of detection which, if considered real, can have strong implications not only on GRB physics but also on the prospects of the detection of VHE emission from the merger of binary neutron star systems, which are established GW emitters. \\
Finally, even if most of the GRBs followed-up by MAGIC do not show any evidence of VHE emission, the upper limits on the VHE flux can still provide valuable information. Especially in the cases where MWL data are available and the redshift is known, upper limits in the MAGIC band can reduce the parameter degeneracy in GRB models. This kind of study is an important reference before the \textit{Cherenkov Telescope Array} is completed, in particular considering that the first \textit{Large Size Telescope} (LST) is currently under commissioning phase and will soon start scientific operations.

\subsection{High Energy Neutrinos}
The origin of ultra high energy cosmic rays (UHECRs) is currently one of the long-standing questions in astroparticle physics. One of the ways to identify the sources of such energetic particles is to search for high energy astrophysical neutrinos, which are a clear sign of hadronic processes involving highly energetic protons.\\
In the current neutrino-detectors panorama, IceCube is the experiment providing most of the scientific information about high energy neutrinos \cite{Aartsenetal2017a}. The discovery of the neutrino diffuse flux \cite{Aartsenetal2016} raises the question of which sources are responsible for the generation of such emission. For this reason, since the production of such energetic neutrinos is related to hadronic processes, which in turn can lead to emission in the HE and VHE bands, the follow-up of neutrino alerts provided by IceCube is crucial to determine their sources. \\
IceCube started providing alerts to gamma-ray facilities since 2012 for clusters of neutrino-like events on a timescale of 3 weeks, the so called gamma-ray follow-up (GFU) program. Starting from 2016, IceCube is distributing automatically alerts on high energy neutrinos thanks to a realtime alert system \cite{Aartsenetal2017b}. Up to 2019, two streams with different selections provided track-like high-energy starting events (HESE) and extremely high-energy (EHE) through-going tracks. In June 2019, these two streams were replaced by the so called Gold and Bronze high-energy neutrino tracks. The number of alerts will increase with respect the old streams providing events with high signal purity. Neutrino alerts are distributed publicly via the \textit{Astrophysical Multimessenger Observatory Network} (AMON) and the GCN. MAGIC promptly joined the effort in the follow-up of neutrino candidate events searching for a correlated VHE emission. Currently the reaction to these alerts is automatic as for GRBs (with different selection criteria). The typical uncertainties on the neutrino reconstructed direction, around \ang{0.2}--\ang{1}, are larger than the MAGIC angular resolution and comparable with MAGIC's field of view. For this reason, a new analysis approach is under development to identify point sources or derive flux upper limits in the localization region provided by IceCube \cite{Fattorinietal2019}.   \\
One remarkable event is the one detected by IceCube on 22nd September 2017 (IC170922A), a neutrino with most probable energy of \SI{290}{\TeV} at 90\% confidence level and median angular resolution of \ang{0.15} \cite{Aartsenetal2018}. The alert was distributed immediately to the astrophysical community\footnote{\url{https://gcn.gsfc.nasa.gov/notices_amon/50579430_130033.amon}} to encourage the follow-up. The neutrino event was found to be consistent with the position of a blazar, TXS 0506+056, known to be a gamma-ray emitter in the HE band. MAGIC performed a first follow-up on the 24th September under nonoptimal conditions without detecting a significant VHE signal. After the report by the \textit{Fermi}-LAT collaboration of an increased flux from TXS 0506+056 (ATel No. 10791), MAGIC performed additional observations of the source for a total of 13 hours between 28th September and 4th October 2017. Integrating the data from these observations, MAGIC data measured a significant VHE emission (at the level of $6.2\sigma$) above \SI{400}{\GeV}: this sets the first time that a gamma-ray flux correlated to an astrophysical neutrino is detected. The emission of such energetic neutrino can be explained if at least part of the total gamma-ray emission has an hadronic origin (see e.g. \cite{Ansoldietal2018,Keivanietal2018,Gaoetal2019}). Therefore, several models involving both leptonic and hadronic processes were proposed to explain both the neutrino and gamma-ray emissions. Given the energy of the neutrino detected by IceCube, the maximum energy of protons accelerated in the jet and leaving the source can reach \SI{e19}{\eV}. Therefore, TXS 0506+056 could contribute as a source of UHECRs \cite{Ansoldietal2018}.

\subsection{Gravitational Waves Counterparts}
The search of EM counterparts to GW candidate events is currently one of the main scientific projects of many facilities, including MAGIC. Indeed, EM emission is expected in systems producing gravitational waves detectable with current ground interferometers. This was demonstrated by the historical detection of the GW signal (GW170817) from a binary neutron star system, whose EM counterpart was seen as a short GRB (GRB\,170817A). The search for counterparts of GW events is tightly connected to the progenitor problem of (short) GRBs. \\
The MAGIC collaboration has been participating to the GW follow-up effort since 2015, when it signed an MoU with the LIGO and Virgo collaboration in order to receive the private alerts on GW candidate events. MAGIC was the first IACT to perform a follow-up of a GW event, GW151226 \cite{Delottoetal2017}. At the time, no procedure was available to perform the automatic follow-up of such events. Therefore the observed targets were chosen taking into account the observations by other facilities and the sky position having the maximum a posteriori probability of being the origin of the GW signal. Moreover, the very large localizations provided only by the LIGO experiments constitute a difficult challenge for IACTs, given their small field of view. For such reason, it is important to have a dedicated strategy for the follow-up of GW events. For the current observational run O3 of LIGO and Virgo, MAGIC set up a (semi-)automatic procedure with different observational cases. \\
For events having a localization region of few tens of square degrees, the sky patch can be covered with a scan (multiple pointings with limited duration) or, given the distance information provided in GW alerts, a correlation with existing catalogs can be performed. If instead a counterpart is found by other instruments and such information is distributed via the GCN Notice channel (e.g.\ GRBs, high energy neutrinos), then the reaction of MAGIC will be automatic. If other channels are used (e.g.\ ATel, GCN Circulars), the counterpart observations can be scheduled easily if the information is distributed during the day or by an expert on duty during the night. Finally, in some cases, like GW170817/GRB\,170817A, delayed EM emission is expected on a timescale of days/weeks. In such a situation, the information provided by other facilities, in particular in the radio, optical and X-ray bands, is crucial to plan the observation schedule. \\
During O3, also KAGRA should join the current network of interferometers. Even better localizations are expected, which highlights the importance of an automatic strategy selecting targets possibly related to the GW events, as soon as possible after the trigger. During O3 the MAGIC strategy will be fully implemented and possibly updated as new information is provided by GW experiments.

\subsection{Fast Radio Bursts}
Fast Radio Bursts (FRBs) are bright flashes of millisecond duration, typically in the \si{\giga\hertz} band with peak fluxes at the Jansky level. Their first discovery by radio telescopes is quite recent and their origin is likely extragalactic. Still, their progenitor sources are unknown. Additionally, some FRBs are seen to exhibit repeating bursts. The first of such reported cases is FRB\,121102 \cite{Spitleretal2014}. Given the repetitive behavior of the source, it was possible to localize it within a dwarf galaxy at $z\sim0.193$. A persistent radio and optical emission was discovered, prompting the search for counterparts in other wavelengths, which was not successful, both for the burst and persistent emissions. In several scenarios explaining the emission from FRB\,121102 and the persistent radio source, VHE emission is expected (see e.g. \cite{Muraseetal2016}). Therefore, MAGIC performed a campaign of simultaneous observations with the radio telescope Arecibo starting in September 2016 \cite{Acciarietal2018}. \\
MAGIC, beside the standard stereo VHE data taking, is able to perform optical observations using the fully modified photosensor-to-readout central pixel of the camera of MAGIC-II. The central pixel can detect the Crab pulsation in the optical band in less than \SI{10}{\second}, achieving a sensitivity of $\sim$\SI{8}{mJy} (13.4 mag) for \SI{1}{\milli\second} optical flashes. \\
During the campaign, Arecibo detected five radio bursts from FRB\,121102. No bright optical pulse was detected in temporal correlation with the MAGIC central pixel. Nevertheless, for the first time it was possible to constrain the burst VHE and optical emission simultaneous to FRBs, which can constrain magnetars models. The limits for the persistent VHE emission instead are comparable to those obtained by VERITAS \cite{Birdetal2017}. \\
Currently the search for FRBs sees many experiments involved, both single dish (Parkes, Arecibo, Effelsberg, FAST) and interferometric radio telescopes (JVLA, WSRT, ASKAP, UTMOST, LOFAR). The recent start of operation of the Canadian Hydrogen Intensity Mapping Experiment (CHIME) is adding more and more events to the FRB sample. In particular, 12 FRBs were detected at low frequencies (down to \SI{400}{\mega\hertz} \cite{Amirietal2019b}), while one event (FRB\,180814.J0422+73) was found to be repeating, with some features similar to those found in FRB\,121102 \cite{Amirietal2019a}. The increase in the number of FRBs, especially those exhibiting a repetitive behavior, is a positive fact for MWL observations, giving the opportunity to have more events to observe simultaneously with radio telescopes. MAGIC in the future will organize other campaigns in order to detect a possible burst VHE/optical emission as performed for FRB\,121102, trying to unveil the progenitor sources and the emission mechanisms of FRBs.

\section{Conclusions and prospects}
Transient sources are excellent laboratories for physics under extreme conditions, providing us with the most energetic events in the Universe. Their observation/follow-up is difficult due to their serendipitous nature: on this aspect. MAGIC is currently one of the telescopes best prepared for transients follow-up, thanks to the hardware, software and technological solutions adopted for this purpose. As far as next searches are concerned, new facilities or upgrades of current ones are expected to come online and provide more events. To name a few, KAGRA and LIGO-India for GW, IceCubGen2 and KM3Net for neutrinos, SKA for FRBs and SVOM and THESEUS for GRBs. Again, a coordinated effort to perform multi-wavelength and multi-messenger observations will be crucial to understand better the origin of such signals. Thanks to continuous technical and analysis improvements MAGIC is currently the leader in the VHE transients search. An envisioned upgrade for transient searches with MAGIC is the use of the so-called \textit{Sum-Trigger-II}, an optional trigger tuned for low energy observations down to \SI{30}{\GeV}. This will have an impact especially for distant sources like GRBs or for intrinsically faint sources in the VHE band.

\section{Acknowledgments}
The authors gratefully acknowledge financial support from the agencies and organizations listed here: \url{https://magic.mpp.mpg.de/acknowledgments_ICRC2019/}.

\end{document}